\documentclass[aps,pra,reprint,superscriptaddress,showpacs]{revtex4-1}
\usepackage{graphicx}
\usepackage{amssymb}
\usepackage{amsmath}
\usepackage{bm}
\usepackage{verbatim}
\usepackage{cancel}
\usepackage[bookmarks=false,colorlinks=true,urlcolor=blue,citecolor=blue,linkcolor=blue]{hyperref}
\usepackage[normalem]{ulem}

\begin{document}

\title{Many-Body Spectral Reflection Symmetry and Protected Infinite-Temperature Degeneracy}

\author{Michael Schecter}
\affiliation{Condensed Matter Theory Center and Joint Quantum Institute, Department of Physics, University of Maryland, College Park, Maryland 20742, USA}

\author{Thomas Iadecola}
\affiliation{Condensed Matter Theory Center and Joint Quantum Institute, Department of Physics, University of Maryland, College Park, Maryland 20742, USA}

\date{\today}

\begin{abstract}
Protected zero modes in quantum physics traditionally arise in the context of ground states of many-body Hamiltonians. Here we study the case where zero modes exist in the center of a reflection-symmetric many-body spectrum, giving rise to the notion of a protected ``infinite-temperature" degeneracy. For a certain class of nonintegrable spin chains, we show that  the number of zero modes is determined by a chiral index that grows exponentially with system size. We propose a dynamical protocol, feasible in ongoing experiments in Rydberg atom quantum simulators, to detect these many-body zero modes and their protecting spectral reflection symmetry. Finally, we consider whether the zero energy states obey the eigenstate thermalization hypothesis, as is expected of states in the middle of the many-body spectrum. We find intriguing differences in their eigenstate properties relative to those of nearby nonzero-energy eigenstates at finite system sizes.
\end{abstract}

\maketitle

\section{Introduction}\label{sec:intro}

Zero modes in quantum physics first came to prominence with the seminal work of Jackiw-Rebbi~\cite{Jackiw76}, Su-Schrieffer-Heeger~\cite{Su79}, and Jackiw-Rossi~\cite{Jackiw81}. They discovered protected zero-energy single-particle states bound to topological defects like solitons in one spatial dimension (1D)~\cite{Jackiw76,Su79} and vortices in 2D~\cite{Jackiw81}.  The robustness of these zero modes was later understood to be guaranteed by an index theorem~\cite{Weinberg81}. Much later, these concepts were generalized to all classes of topological insulators (TIs), which generically have protected zero modes at topological defects of various codimensions, including spatial boundaries~\cite{Teo10}.

Protected zero modes also manifest themselves in supersymmetric (SUSY) lattice models~\cite{Nicolai76,Witten81}. Unlike their counterparts in TIs, SUSY zero modes are many-body entities whose existence does not require spatial boundaries or defects. However, their robustness is also guaranteed by an index theorem due to Witten~\cite{Witten82}. In both cases, zero modes arise in the context of \emph{ground states} of many-body Hamiltonians and are therefore relevant at low energies; in SUSY, zero-energy states must be ground states, while in TIs the zero-energy single-particle states sit atop a filled Fermi sea of negative-energy states. 

In this paper, we explore a class of quantum spin systems that host symmetry-protected zero modes at \emph{finite energy densities} above the ground state. They are protected by a reflection symmetry of the energy spectrum of the many-body Hamiltonian $H$, generated by an operator $\mathcal C$ satisfying $\{\mathcal C,H\}=0$, which pins the zero modes to the center of the spectrum. We classify these zero modes by a symmetry-resolved index and propose a dynamical protocol that allows one to measure the number of zero modes systematically. We exemplify these results in a nonintegrable spin system motivated by the mixed-field Ising chain near the saturation field, which can be simulated using Rydberg atoms in optical lattices \cite{Bernien17,Guardado-Sanchez17,Lienhard17}.

The existence of spectral reflection symmetry implies that every eigenstate $|E\rangle$ of $H$ has a chiral partner $\mathcal C |E\rangle=|{\rm -}E\rangle$. Zero modes of $H$, if they exist, are unique among eigenstates of $H$ because they can be chosen to diagonalize $\mathcal C$ and acquire definite chiral charge. As a result, one may define an index $W={\rm tr}\left(\mathcal C\, e^{-\beta H}\right)$ that lower-bounds the number of zero modes $N_0\geq |W|$, similar to the Witten index of SUSY (here, $\beta$ is the inverse temperature). 
When the Hamiltonian has a symmetry $\mathcal S$ that commutes with $\mathcal C$, one can define an index
\begin{align}
\label{eq: general index}
W_{\mathcal S}={\rm tr}\left( P_{\mathcal S}\, \mathcal C\, e^{-\beta H}\right)
\end{align}
for each symmetry sector of $\mathcal S$ using the projector $P_{\mathcal S}$. 
The number of zero modes thus obeys a much stronger bound in this case: $N_0\geq {\rm tr}_{\mathcal S}|W_{\mathcal S}|$. In this work, we show that a striking scenario arises when the total charges of $\mathcal C$ and $\mathcal S$ in the zero-mode manifold are $\mathcal O(1)$, while $ {\rm tr}_{\mathcal S}|W_{\mathcal S}|\gg 1$. This implies that the intertwining of $\mathcal C$ and $\mathcal S$ in the zero-mode manifold can lead to a dramatic enhancement of the number of zero modes. As we discuss later, this intertwining of symmetries in the zero-mode manifold can also be exploited for their detection.

Here we focus on point-group symmetries and show that they can lead to \emph{exponential} growth of the number of zero modes with system size $L$, similar to superfrustrated SUSY models \cite{Fendley05,Nicolai76,vanEerten05,Huijse08,Huijse12,Katsura17}.
The simplest example is the paramagnet with Hamiltonian
\begin{align}
\label{eq: H_para}
H_{\rm para}=\sum_i X_i,
\end{align} where $X_i,Y_i,Z_i$ are Pauli operators on sites $i=1,\dots,L$ of a lattice with point-group symmetry $\mathcal S$. The spectral-reflection operator
\begin{align}
\label{eq: C def}
\mathcal C=\prod_i Z_i
\end{align}
measures the parity of the number of ``down" spins. Constructing zero modes of $H_{\rm para}$ is straightforward for even $L$: align half the spins parallel to $X$, and the other half antiparallel. The number of zero modes, $N_0=\binom{L}{L/2}\sim 2^L$, grows exponentially with system size.

At first glance, this dramatic growth of the number of zero modes with $L$ is a trivial consequence of the integrability of the paramagnet. However, the existence of an exponentially large index $W_{\mathcal S}$ guarantees that it is not. Rather, exponentially many zero modes of the paramagnet persist in the presence of \emph{arbitrary} perturbations that preserve spectral reflection symmetry and the point-group symmetry $\mathcal S$. For example, one can add to the Hamiltonian $H_{\rm para}$ in Eq.~\eqref{eq: H_para} a set of terms that anticommute with $\mathcal C$ and commute with $\mathcal S$, such as
\begin{align}
\label{eq: general deltaH}
\begin{split}
\delta H
=&
\sum_{\langle ij \rangle}\! \left(a_{0,ij}\, Z_{i}X_{j}+a_{1,ij}\, X_{i}Z_{j}+\dots\right)\\
&+\sum_{\langle ijk \rangle}\!\left(a_{2,ijk}\, Z_{i}X_{j}Z_{k}+a_{3,ijk}\, X_{i}X_{j}X_{k}+\dots\right)\\
&+\dots,
\end{split}
\end{align}
where $\langle\ \cdot\ \rangle$ denotes that the enclosed indices label nearest-neighbor sites.  Here, in order to ensure $\{\mathcal C,\delta H\}=0$, the only allowed terms are those for which the total number of operators $O_i=X_i$ or $Y_i$ is odd. Furthermore, the coefficient of each term must be chosen such that the point group symmetry $\mathcal S$ is maintained. For the case of a 1D lattice where the point group symmetry $\mathcal S$ is given by spatial inversion symmetry $\mathcal I$, the symmetry-resolved index \eqref{eq: general index} is given by
\begin{align}
\label{eq: general 1D inversion index}
W_\pm=\pm 2^{L/2-1}\indent \text{for even $L$},
\end{align}
where $\pm$ label the eigenvalues $\pm1$ of $\mathcal I$. One thus has $N_0\geq 2^{L/2}$ zero modes, despite the presence of strong integrability-breaking perturbations. Moreover, the zero modes are even robust to breaking $\mathcal I$ as long as $\{\mathcal{C I},H\}=0$, in which case $N_0\geq \left|{\rm tr}\left(\mathcal{CI}\, e^{-\beta H}\right)\right|=2^{L/2}$. In other words, as long as one can define an appropriate spectral reflection symmetry, these zero modes persist.

Like the Witten index, the indices $W_{\mathcal S}$ are well-defined at finite temperature. Unlike the Witten index, however, $W_{\mathcal S}$ is trivially zero at zero temperature, since the density operator $e^{-\beta H}$ becomes
a projector onto the ground state in the limit $\beta\to\infty$. The latter fact suggests that physical signatures of the spectral reflection symmetry and zero modes become important only at high temperatures, or in the far-from-equilibrium dynamics of the system. We discuss the physical consequences of the exponentially large zero-mode manifold below in Sec.~\ref{sec:signatures}. 

The rest of the paper is organized as follows. In Sec.~\ref{sec:model} we review a model relevant for ongoing experiments studying Rydberg-atom arrays and show how the low-energy sector of the Hilbert space asymptotically acquires the spectral-reflection symmetry. In Sec.~\ref{sec:indices} we introduce and calculate the symmetry-resolved chiral indices for the low-energy, projected model. In Sec.~\ref{sec:signatures} we discuss how to detect the presence of zero modes and their effect on the Loschmidt echo dynamics of experimentally preparable product states. We also consider the question of whether or not the zero modes obey the eigenstate thermalization hypothesis. Conclusions are summarized in Sec.~\ref{sec:conclusion}.

\section{Model}\label{sec:model}

In this paper we focus on a model that is relevant to ongoing experiments studying arrays of Rydberg atoms \cite{Bernien17,Lienhard17},
namely the mixed-field Ising chain with the Hamiltonian
\begin{equation}
\label{eq:H}
H=\sum_{i<j}V_{ij}\, Z_i Z_j+\sum_i\left(h_x\, X_i+h_z\, Z_i\right).
\end{equation}
Here, $h_z,h_x$ are the longitudinal and transverse fields, and $V_{ij}$ is a repulsive (antiferromagnetic) interaction. This system can be simulated using Rydberg atoms in an optical lattice, where $V_{ij}$ arises due to van der Waals coupling between atoms and therefore decays rapidly with $|i-j|$. In the optical tweezer arrays
of Refs.~\cite{Bernien17,Lienhard17}, and in the quantum gas microscope of Ref.~\cite{Guardado-Sanchez17}, the interatomic spacing can be tuned, allowing one to selectively truncate to nearest or next-nearest neighbor coupling. Unless otherwise specified, we restrict ourselves to the nearest-neighbor case and denote the nearest-neighbor coupling $V_{ii+1}\equiv V_1$.

In the limit  $h_{x}\ll V_1$ and near the saturation field $h_z\sim 2V_1$, the low-energy eigenstates of Eq.~\eqref{eq:H} are linear combinations of $Z_i$ eigenstates in which no two neighboring spins point ``up." This means that the effective low-energy Hamiltonian $\tilde H$ can be written (up to an overall energy shift) using projectors as~\cite{Lesanovsky12,Bernien17}
\begin{equation}
\label{eq:H1}
\tilde H=\sum_i\left(h_x\,\tilde{X}_i+\Delta\, \tilde{Z}_i\right),
\end{equation}
where $\Delta=h_z-2V_1$, $\tilde{O}_i=O_i\prod_{j\in\mathrm{nn}(i)}P_{j}$ and $P_i=(1-Z_i)/2$ is the local projector onto spin-down. For $\Delta=0$, the Hamiltonian $\tilde H$ acquires a spectral reflection symmetry generated by $\mathcal{C}=\prod_i Z_i$, just as for the paramagnet discussed above. Unlike the paramagnet, however, the Hamiltonian~\eqref{eq:H1} is strongly interacting and nonintegrable due to the low-energy constraint imposed on the Hilbert space. It is straightforward to generalize Eq.~\eqref{eq:H1} to higher-dimensional bipartite lattices, where the saturation field is $h_z=z_c V_1$ with $z_c$ the coordination number.

\section{Symmetry-resolved Chiral Indices}\label{sec:indices}

As pointed out in Ref.~\cite{Lesanovsky12}, $\tilde H$ (sometimes called the ``Fibonacci chain") equivalently describes a system of Fibonacci anyons \cite{Feiguin07,Trebst08,Chandran16,Chen17} whose Hilbert space dimension $\mathcal D(L)=F_{L+2}\sim \varphi^L$, where $F_i$ are the Fibonacci numbers (with $F_1=1$ and $F_{i+1}=F_{i}+F_{i-1}$, which yields the famous sequence $1,1,2,3,5,\dots$) and $\varphi=1.618...$ is the golden ratio. A recent experiment using Rydberg atoms~\cite{Bernien17} has shown that this system exhibits peculiar quench dynamics in the form of  persistent oscillations that last long after the natural timescale of $\tilde H$, $1/h_x$. In Ref.~\cite{Turner17} this phenomenon was attributed to ``scarring" of the many-body wavefunction in analogy to single-particle quantum chaos. The authors of Refs.~\cite{Chandran17,Turner17} also pointed out the existence of an exponentially large number of zero modes of $\tilde H$ that are sensitive to  inversion symmetry.

Here we see that, in the presence of the $\mathbb Z_2$ inversion symmetry $\mathcal I=\mathcal S$,
such zero modes are guaranteed by an index,
\begin{equation}
\label{eq:index}
W_\pm={\rm tr}\left(\frac{1\pm\mathcal{I}}{2}\, \mathcal{C}\, e^{-\beta H}\right),
\end{equation}
where the trace is taken over the \emph{constrained} Hilbert space. The total number of zero modes satisfies $N_0\geq |W_+|+|W_-|$. For the Fibonacci chain, the indices can be computed explicitly; for open boundary conditions, they are given by
\begin{equation}
\label{eq:indices}
W_\pm = \begin{cases}
\frac{-a(L)\pm F_{L/2+1}}{2} & L\,\,{\rm even}\\
\frac{a(L)\mp F_{(L-1)/2}}{2} & L\,\,\,{\rm odd},
\end{cases}
\end{equation}
where $a(L)=\frac{1}{2}(-1)^{\left\lfloor(L-2)/3\right\rfloor}+\frac{1}{2}(-1)^{\left\lfloor(L-1)/3\right\rfloor}$ is related to ${\rm tr}\,\mathcal C$ and $\left\lfloor \cdot \right\rfloor$ is the integer part. Since the sign of $W_\pm$ is determined by the chiral charge of the zero modes, we find that for even $L$ the inversion even (odd) zero modes have positive (negative) chiral charge,  while for odd $L$  inversion even (odd) zero modes have negative (positive) chiral charge. As we show in Sec.~\ref{sec:signatures}, this intertwining of chiral charge and inversion symmetry eigenvalues in the zero-mode manifold is an important feature that can be exploited to measure the zero-mode count. The total number of zero modes of Hamiltonian~\eqref{eq:H1} at $\Delta=0$ in fact saturates the bound $N_0\geq|W_+|+|W_-|$, namely
\begin{equation}
\label{eq:zero-modes}
N_0=\begin{cases}
 F_{L/2+1} & L\,\,{\rm even},\\
 F_{(L-1)/2} & L\,\,\,{\rm odd},
\end{cases}
\end{equation}
in agreement with Refs.~\cite{Chandran17,Turner17}.
For large $L$, this implies that $N_0(L)\sim \varphi^{L/2}$, and thus $N_0(L)\sim\sqrt{\mathcal D(L)}$.

One can readily generalize the results (\ref{eq:indices})--(\ref{eq:zero-modes}) to the case where the $k$th nearest neighbor coupling $V_k$ exceeds $h_x$. As shown in Ref.~\cite{Bernien17}, this leads to a sequence of $\mathbb Z_k$ symmetry-broken ground states. The low-energy subspaces can be obtained as before by dressing operators with projectors that eliminate Rydberg excitations (``up spins") within a radius of $k$ sites: $\prod_{1\leq j\leq k}P_{i-j}P_{i+j}$. The dimension of the constrained Hilbert space can be computed recursively in terms of the system length. For an open chain, the constrained Hilbert-space dimension at system size $L$ obeys
\begin{align}
\mathcal D_k (L)
=
\begin{cases}
\mathcal D_k(L-1)+\mathcal D_k(L-k-1) & L>k+1,\\ 
\mathcal D_k(L)=L+1 & L\leq k+1
\end{cases}.
\end{align}
In terms of this sequence, one can compute the index ${\rm tr}\left(\mathcal{C I}e^{-\beta H}\right)$ explicitly, leading to the lower bound
\begin{equation}
\label{eq:indices2}
N_{k,0} \geq \begin{cases}
\mathcal D_k\left(\frac{L}{2}-\lfloor \frac{k+1}{2}\rfloor\right) & L\,\,{\rm even},\\
\mathcal D_k\left(\frac{L-1}{2}-\lfloor\frac{k}{2}\rfloor\right)- \mathcal D_k\left(\frac{L-1}{2}-k\right) & L\,\,\,{\rm odd}.
\end{cases}
\end{equation}
The above results reduce to those of the previous paragraph in the case $k=1$, where $\mathcal D_1(L)\equiv\mathcal D(L)$. For large $L$, the constrained Hilbert space dimension grows exponentially, $\mathcal D_k(L)\sim \alpha^L$, where the base $\alpha$ is the positive root of $\alpha^{k+1}-\alpha^k-1=0$. From Eq.~\eqref{eq:indices2} we see that for all $k$, $N_{k,0}(L)\gtrsim \sqrt{\mathcal D_k(L)}$ for $L\gg k$.

That $N_0$ scales with the square root of the total Hilbert space dimension is a generic consequence of the fact that $\mathcal I$ is a $\mathbb Z_2$ symmetry and the fact that spins on different sites commute. Since the first term in $W_\pm$, $\frac{1}{2}{\rm tr}\left(\mathcal{C}e^{-\beta H}\right)$, is $\mathcal O(1)$, the total number of zero modes is actually bounded by ${\rm tr}\left(\mathcal{C I}e^{-\beta H}\right)=\sum_{n\in\{n\}_\mathcal{I}}\mathcal{C}_n$, where $\{n\}_\mathcal{I}$ denotes the set of inversion-invariant product states of $Z_i$.  In a system with even $L$ it is clear that every inversion-invariant state has $\mathcal{C}=1$ and $\mathcal I=1$. For odd $L$, exponentially more inversion-invariant states have $C=-1$ due to the Hilbert space constraints. This leads to ${\rm tr}(\mathcal{C I})<0$ and the opposite pairing (compared to even $L$) of $\mathcal C$ and $\mathcal I$ eigenvalues in the zero-mode manifold. In either case, the $\mathbb Z_2$ inversion symmetry effectively halves the number of degrees of freedom in the trace, thus giving $N_0\sim \sqrt{\mathcal D}$. This square-root scaling of the number of zero modes is also present in the example of the paramagnet \eqref{eq: H_para} perturbed by the generic inversion- and chiral-symmetric interactions \eqref{eq: general deltaH}, c.f.~Eq.~\eqref{eq: general 1D inversion index}.

\section{Dynamical Signatures of Spectral Reflection Symmetry and Many-Body Zero Modes}\label{sec:signatures}

In this Section we discuss possible physical consequences of the existence of many-body zero modes protected by a spectral reflection symmetry. In Sec.~\ref{sec:counting}, we show how to detect the presence of zero modes and their chiral index by studying the late-time dynamics of the chiral charge $\mathcal C(t)$ when the system is initialized in a $Z_i$ product state. In Sec.~\ref{sec:Loschmidt}, we make the connection between the dynamics of the chiral charge and the Loschmidt echo of arbitrary $Z_i$ product states, showing how the presence of zero modes drastically enhances the echo response. Finally, in Sec.~\ref{sec:ETH}, we discuss the notion of eigenstate thermalization within the exponentially large zero-mode manifold, and present numerical evidence supporting a modified version of ETH for local operators evaluated in the manifold.

\subsection{Zero-Mode Index from Chiral Charge Dynamics}\label{sec:counting}

\begin{figure}[t!]
\includegraphics[width=\columnwidth]{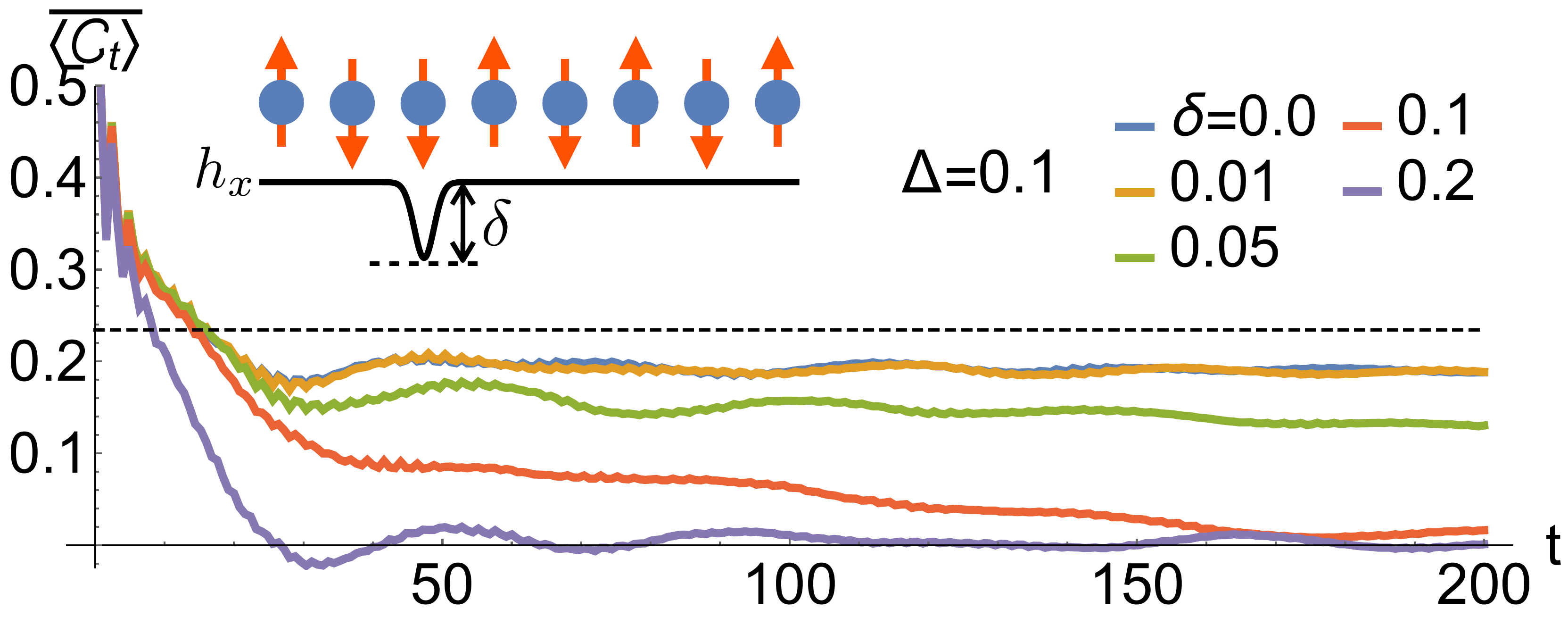}
\caption{(Color online)
Dynamics of the moving average $\overline{\langle\mathcal C_t\rangle}=\int_0^t \frac{dt^\prime}{t}\langle\psi| \mathcal C (t^\prime)|\psi\rangle$,
where $|\psi\rangle=|\uparrow\downarrow\dots\rangle$ is the N\'eel state, for $L=8$.  Time $t$ is measured in units of $h^{-1}_{x}$, and energy is measured in units of $h_x$. We set $\Delta=0.1$ to slightly break the spectral-reflection symmetry.
The strong sensitivity of $\overline{\langle\mathcal C_t\rangle}$ to variations of the $\mathcal I$-breaking
energy scale $\delta$ is indirect evidence of the symmetry-protected zero modes, see discussion after Eq.~\eqref{eq:obs}. ($\delta$ is defined pictorially in the inset as a local substitution $h_x\to h_x-\delta$ on a single off-centered site.) For $\delta\gtrsim \Delta$ the late-time value of $\overline{\langle\mathcal C_t\rangle}$ approaches zero rapidly. The dashed line indicates the infinite-time value $\overline{\langle\mathcal C_\infty\rangle}$
for $\Delta=\delta=0$.
}
\label{fig:C_avg}
\end{figure}

The most direct signature of the many-body zero modes arises from studying the dynamics of the average chiral charge $\langle\mathcal C(t)\rangle$. Below, we show that this quantity serves as a sensitive indicator for zero modes. We take the initial states to be arbitrary (but constrained) $Z_i$ product
 states, which are readily preparable experimentally. Chiral charge can be measured by simply counting the number of ``down" spins in the final state, $N_{\downarrow}=\sum_i (1-Z_i)/2$, giving $\mathcal C=(-1)^{N_{\downarrow}}$. Denoting the initial state by $|\psi\rangle$, the time-averaged chiral charge is given by
 \begin{align}
 \overline{\langle\mathcal C_t\rangle}_\psi\equiv\int_0^t \frac{dt^\prime}{t}\langle\psi| \mathcal C (t^\prime)|\psi\rangle.
 \end{align}
 An example is shown in Fig.~\ref{fig:C_avg}, where $|\psi\rangle$ is the N\'{e}el state. In the presence of spectral reflection symmetry, the late-time average $\lim_{t\to\infty}\overline{\langle\mathcal C_t\rangle}_\psi\equiv\overline{\langle\mathcal C_\infty\rangle}_\psi$ can be written in the eigenbasis of $H$ as 
\begin{equation}
\label{eq:obs}
\overline{\langle\mathcal C_\infty\rangle}_\psi=\sum_{ E=0} \langle E|\psi\rangle\langle \psi |E\rangle\,  \mathcal C_{E},
\end{equation}
where $\mathcal C_E=\langle E|\mathcal C|E\rangle$.
We notice that in Eq.~\eqref{eq:obs} only the zero modes, which have definite chiral charge $\mathcal C_E=\pm1$, contribute to the long-time expectation value. As a result, one can reconstruct the index ${\rm tr}\,\mathcal C$ by summing the late-time average Eq.~\eqref{eq:obs} over the complete set of initial product states: ${\rm tr}\,\mathcal C=\sum_\psi\overline{\langle\mathcal C_\infty\rangle}_\psi$. However, this index is $\mathcal O(1)$ and does not capture the exponentially large number of zero modes.

\begin{figure}[t!]
\includegraphics[width=0.5\textwidth]{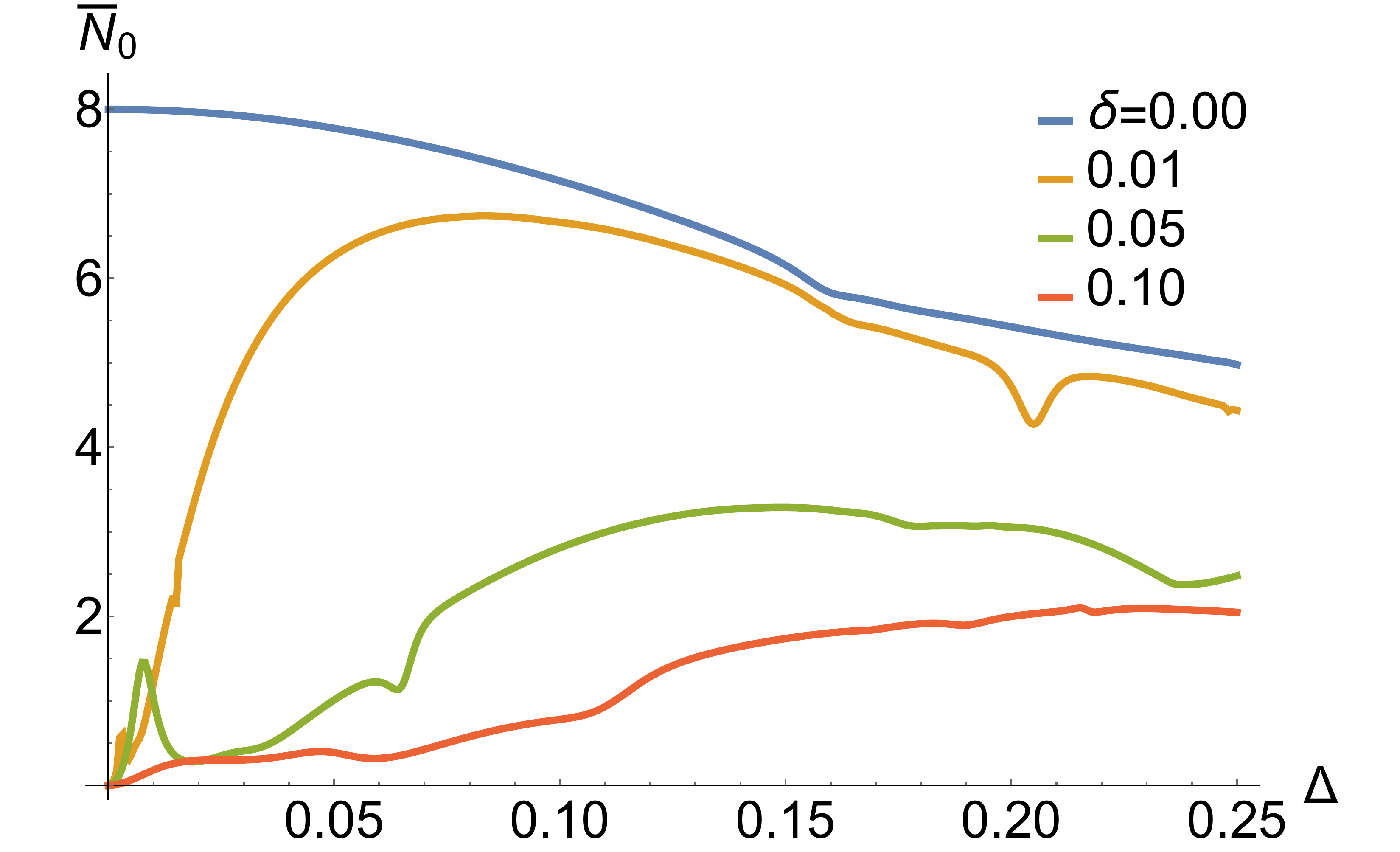}
\caption{
(Color online)
The quantity $\overline N_{0}$, Eq.~\eqref{eq:N0}, at system size $L=10$ as a function of the $\mathcal C$-breaking energy scale
$\Delta$ for different values of the $\mathcal I$-breaking energy scale $\delta$ (see inset of Fig.~\ref{fig:C_avg}). The initial state
$|\psi\rangle$ is again the N\'eel state.
When $\delta=0$ and $\mathcal I$ is preserved, $\overline N_{0}=N_0=8$ at $\Delta=0$, as expected,
and decreases smoothly for $\Delta>0$. When $\mathcal I$ is broken, $N_0=0$ and $\overline N_{0}$ decreases sharply to zero.
}
\label{fig:N0_C-inversion_breaking}
\end{figure}

Naively, it appears that in order to reconstruct the indices $W_\pm$ one must make a challenging simultaneous measurement of $\mathcal C$ and $\mathcal I$. However, one can show that measuring $\mathcal I$ is not necessary due to the pairing of $\mathcal C$ and $\mathcal I$ eigenvalues in the zero-mode manifold [see discussion below Eq.~\eqref{eq:indices}]. If one can group the zero modes by their chiral charge, they will inevitably be grouped by their inversion eigenvalue as well.

It turns out that choosing the initial states $|\psi\rangle$ to be $Z_i$ product states automatically groups the set of zero modes entering Eq.~\eqref{eq:obs} by their chiral charge. Since each such product state has a definite chiral charge $\mathcal C_\psi=\pm1$, it can only project onto zero modes with the same chiral charge. As a result, the indices $W_\pm$ can be obtained simply by restricting the summation in \eqref{eq:obs} to run over the set of initial product states with a particular chiral charge:
\begin{equation}
\label{eq:Wpm}
W_\pm =(-1)^L\!\! \sum_{\psi, \mathcal C_\psi=\pm1} \overline{\langle\mathcal C_\infty\rangle}_\psi,
\end{equation}
where $(-1)^L$ accounts for the fact that chiral-charge and inversion eigenvalues are paired oppositely for even/odd $L$. The total number of zero modes for the open Fibonacci chain at $\Delta=0$ is given by 
$N_0=\overline N_0$ where
\begin{equation}
\label{eq:N0}
\overline N_0=\left|\sum_{\psi, \mathcal C_\psi=1}\overline{\langle\mathcal C_\infty\rangle}_\psi \right|+\left|\sum_{\psi, \mathcal C_\psi=-1}\overline{\langle\mathcal C_\infty\rangle}_\psi \right|.
\end{equation}

It is important to stress that the quantity
$\overline N_0$
is \textit{only} strictly quantized to $N_0$ in the presence of spectral reflection symmetry, and is highly sensitive to the presence of inversion symmetry. This is seen in the $\Delta\to0$ limit of Fig.~\ref{fig:N0_C-inversion_breaking} where
$\overline N_0\to 0$
rapidly upon even slightly breaking $\mathcal I$.
[We break $\mathcal I$ by changing $h_x\to h_x-\delta$ on a non-centered site (see inset of Fig.~\ref{fig:C_avg}).]  It is also evident in
the dynamics of $\langle\mathcal C(t)\rangle_\psi$,
as seen in Fig.~\ref{fig:C_avg}.  The sensitivity to inversion breaking is smoothed out in the presence of a weak reflection-symmetry-breaking perturbation $\Delta$. While a finite $\Delta$ also abruptly changes the zero-mode count, we see from Fig.~\ref{fig:N0_C-inversion_breaking} that
$\overline N_0$
changes smoothly with $\Delta$, becoming non-quantized. This is due to the fact that once reflection symmetry is broken, states with $E\neq 0$ also contribute to $\overline{\langle\mathcal C_\infty\rangle}_\psi$ and Eq.~\eqref{eq:obs} no longer holds. This will also be true for the mixed-field Ising model \eqref{eq:H} where spectral-reflection symmetry breaking terms arise at order $h_x^2/V_1$~\cite{Lesanovsky12} even for $h_z=2V_1$. In the experiment of Ref.~\cite{Bernien17} the small parameter $h_x/V_1\approx 0.04$ indicates that the spectral-reflection symmetry breaking due to virtual processes is a small perturbation compared to the direct symmetry breaking term $\propto \Delta$, when $\Delta/h_x\geq 0.04$.

The results shown in Figs.~\ref{fig:C_avg}--\ref{fig:N0_C-inversion_breaking} indicate that
$\overline N_0$
is far more sensitive to inversion symmetry breaking than spectral-reflection symmetry breaking. In the presence of both symmetry-breaking perturbations, there is a crossover between the two limits of exponentially large
$ \overline N_0$ and
$ \overline N_0\sim \mathcal O(1)$
when the perturbation strengths become comparable. 

We note that although we have shown that the zero-mode count can be measured, in principle, by summing over all initial product states with fixed chiral charge, there is a practical limit to this protocol: the requisite number of initial product states one must prepare in an experiment grows exponentially with system size.
As discussed in the Appendix, it is possible to overcome this drawback via a random sampling of $\overline{\langle\mathcal{C}_\infty\rangle}_\psi$ over initial states $\psi$.

\subsection{Loschmidt Echo}\label{sec:Loschmidt}

\begin{figure}[t!]
\includegraphics[width=\columnwidth]{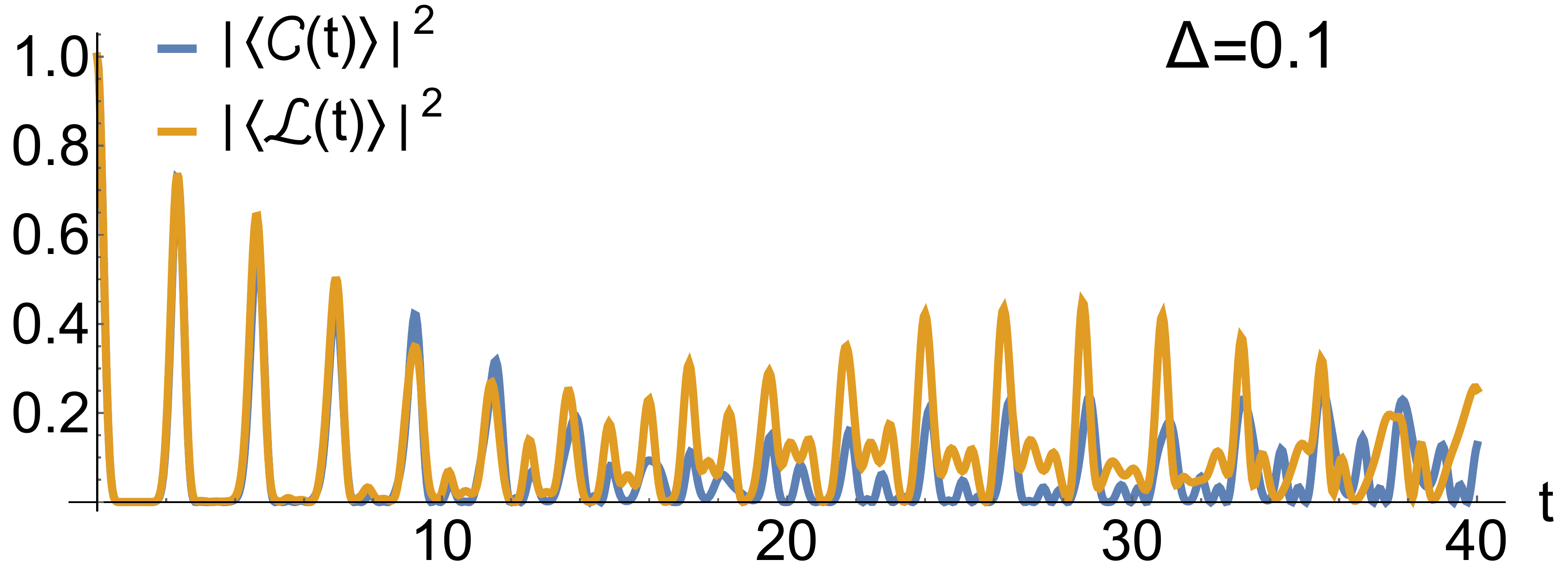}
\caption{(Color online)
Temporal correlation between $\langle \mathcal C (t)\rangle$
and the Loschmidt echo $\mathcal L(t)=\langle e^{-2iHt}\rangle$ in the Fibonacci chain \eqref{eq:H1}
at $L=8$ starting from the N\'{e}el state $\mid\uparrow\downarrow\uparrow\downarrow\uparrow\downarrow\uparrow\downarrow\rangle$. We choose $\Delta=0.1$ to weakly break the spectral reflection symmetry. The two quantities exhibit near-perfect correlation out to a
time $t\sim1/\Delta$, where $t$ is measured in units of $h^{-1}_{x}$.
}
\label{fig:C_Loschmidt_comparison}
\end{figure}

We now discuss the dynamics of the Loschmidt echo, $\mathcal L(t)=\langle e^{iH_2 t}e^{-iH_1t}\rangle$, which has been extensively studied in other models both in the context of quantum information and quantum chaotic systems \cite{Gorin06,Goussev16}. We focus on the special case where the backward evolution is dictated by the Hamiltonian $H_2 = - H_1\equiv H$, in which case it is readily seen that the Loschmidt echo dynamics of any eigenstate of $\mathcal C$ (spanned by $Z_i$ product states) exhibits perfect time-correlation with the expectation value of the chiral charge $\left|\langle \mathcal C (t)\rangle\right|=\left|\langle e^{-2iHt}\rangle\right|$ in the presence of spectral reflection symmetry. This follows from the fact that $\mathcal C$ also acts as a ``time-reflection" operator \cite{Iadecola17} for $Z_i$ product states (not to be confused with time-reversal $\mathcal T$), which sends $t\to-t$ without complex conjugation. When the spectral reflection symmetry is weakly broken, the temporal correlations persist up to a time of order the inverse strength of the perturbation, allowing one to measure the symmetry breaking directly, as shown in Fig.~\ref{fig:C_Loschmidt_comparison}. In the system described by Eq.~\eqref{eq:H} at $h_z=2V_1$, this measurement could allow one to detect the degree of spectral reflection symmetry breaking due to virtual processes involving higher energy states with nearest-neighbor ``spin-up" defects.

\begin{figure}[t!]
\includegraphics[width=\columnwidth]{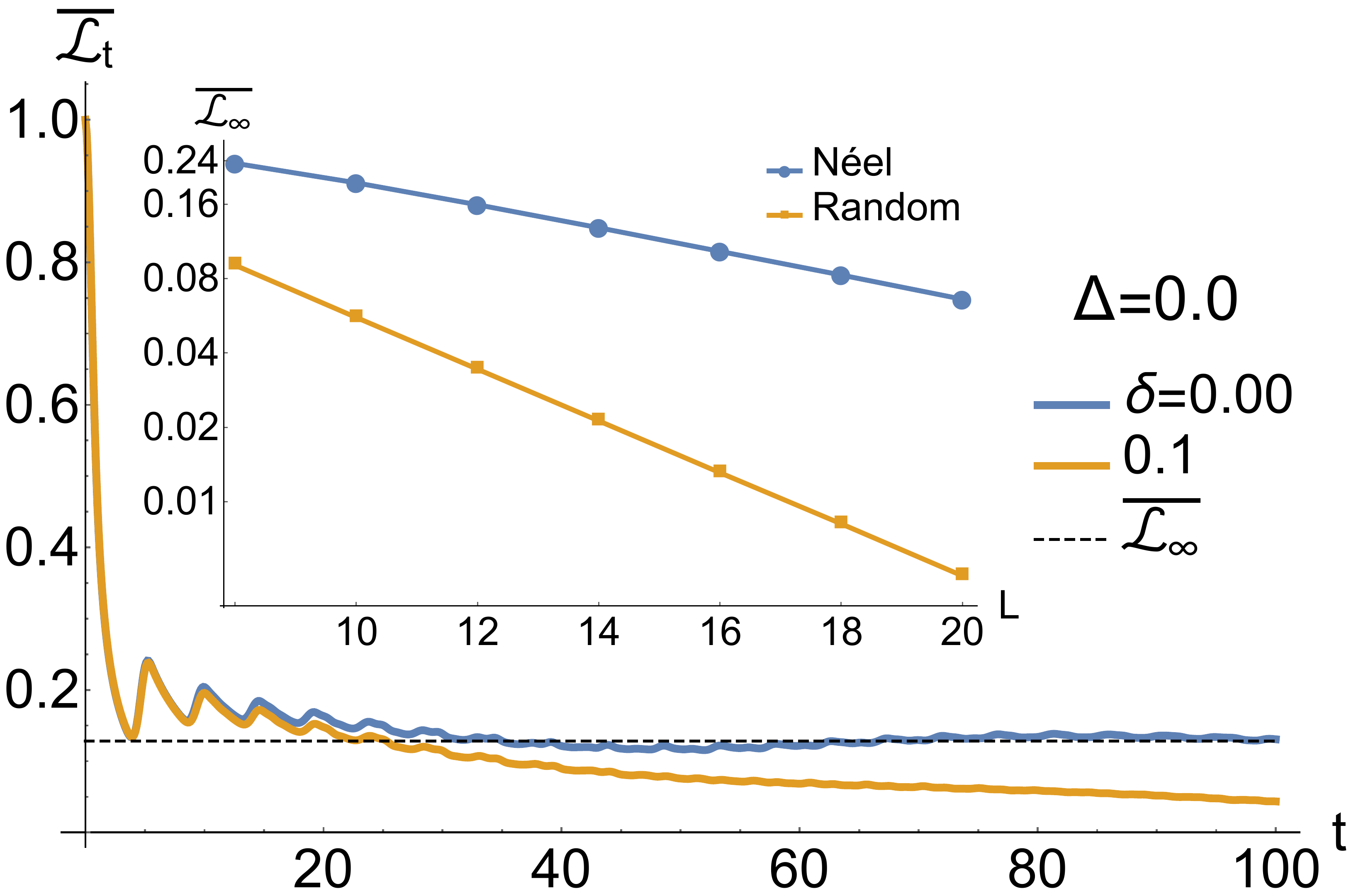}
\caption{(Color online)
Moving average of the Loschmidt echo starting from the N\'{e}el state for $L=14$. The infinite-time value (dashed black line) is finite only in the presence of zero modes. The decaying curve occurs when the zero modes are lifted by breaking inversion symmetry, which is done by reducing $h_x$ on site $3$ by 10\%. The inset shows the infinite-time value $\overline{\mathcal{L}_\infty}$ as a function of system size $L$ in the presence of zero modes.  A logarithmic scale is used on the vertical axis, so that a straight line indicates exponential scaling. The scarring leads to a significant enhancement compared to a random state, $\overline{\mathcal L_\infty}= N_0/\mathcal{D}$ (yellow).
}
\label{fig:Loschmidt_avg}
\end{figure}

Importantly, since each eigenstate of $\mathcal C$ has exactly zero average energy, $\langle H\rangle=0$, they are nominally ``infinite-temperature" states. This implies that, generically, late-time observables initiated in $\mathcal C$ eigenstates should be controlled by energy eigenstates in the middle of the many-body spectrum, where the zero modes are pinned. As we show below, one of the defining features of the presence of such zero modes is the relatively large residual value of the time-averaged Loschmidt echo at late times, $\overline{\mathcal L_{\infty}}\equiv\overline{\mathcal L (t\to\infty)}$.

Any finite residual value of $\overline{\mathcal L_{\infty}}$ is generically unexpected because the infinite-time average washes out any oscillating contribution. However, in the presence of zero modes the time-averaged Loschmidt echo at late times becomes 
\begin{equation}\label{eq:Loschmidt}
\overline{\mathcal L_{\infty}}=\sum_{E=0} \langle E|\psi\rangle\langle \psi |E\rangle.
\end{equation}
If the initial state $|\psi\rangle$ is an eigenstate of $\mathcal C$, and thus nominally an infinite-temperature state with respect to $H$ when $\{\mathcal C,H\}=0$, we expect its overlap with any eigenstate to scale with Hilbert space dimension as
\begin{align}
\langle E|\psi\rangle \propto 1/\sqrt{\mathcal D}.
\end{align}
Since the dimension of the zero-mode manifold scales with $\sqrt{\mathcal D}$ and therefore increases exponentially with system size, it follows that the average Loschmidt echo becomes $\overline{\mathcal L_{\infty}}\propto 1/\sqrt{\mathcal D}$, whereas it would be zero in the case without zero modes.  We show this behavior for the Fibonacci chain \eqref{eq:H1} in Fig.~\ref{fig:Loschmidt_avg} for cases with and without zero modes. In contrast to the dynamics of chiral charge discussed in Sec.~\ref{sec:counting},  the late-time value of the Loschmidt echo only depends on the presence of zero modes and is thus roughly equally sensitive to the breaking of $\mathcal C$ as it is to breaking $\mathcal I$.

It is also worth mentioning that the Fourier transform of the Loschmidt echo, $\tilde{\mathcal{L}}(\omega)$, determines the statistics of work done on a system after a quantum quench~\cite{Goussev16}. Equation~\eqref{eq:Loschmidt} then represents the amplitude of a quench to perform \emph{zero} work on the system, which can occur only in the presence of zero modes.

Another intriguing aspect of the Fibonacci chain worth returning to is the ``scarring" of the many-body eigenstates~\cite{Turner17}. The scarring leads to a significant enhancement of the projection of the N\'{e}el states onto the zero-mode manifold (as well as finite-energy scarred bands). The projection of an initial state onto the zero modes is given by Eq.~\eqref{eq:Loschmidt} and can therefore be measured by the late-time dynamics of $\mathcal L(t)$. In the inset of Fig.~\ref{fig:Loschmidt_avg}, we show the infinite-time value of the Loschmidt echo for the Fibonacci chain as a function of system size $L$. In contrast to the N\'{e}el states, a random (infinite temperature) state projects onto the zero-mode manifold with a weight $N_0/\mathcal D\sim1/\sqrt{\mathcal D}$. As indicated in the inset of Fig.~\ref{fig:Loschmidt_avg},  the presence of scarred zero modes in certain initial states thus provides a substantial increase in the late-time value of the Loschmidt echo for moderately large systems.

\subsection{Zero-Mode Eigenstate Thermalization}\label{sec:ETH}

We now turn to the question of the dynamics of generic local observables $\mathcal{O}$ in the Fibonacci chain, focusing in particular on the role played by the presence of zero modes. To facilitate the discussion it is useful to introduce the  \emph{eigenstate thermalization hypothesis} (ETH) \cite{Deutsch91,Srednicki94,D'Alessio16,Mori17}, first developed to explain how closed quantum systems approach thermal equilibrium despite unitary time evolution. Formally, one may say that a closed quantum system thermalizes when the reduced density matrix of an arbitrary finite subsystem approaches the Boltzmann/Gibbs thermal density matrix at late times after a quantum quench. The temperature of the thermal ensemble is determined by the global energy density of the initial state. However, the equivalence between the grand-canonical and microcanonical ensembles implies that \emph{any} single eigenstate $|E\rangle$ with the correct energy density~\cite{D'Alessio16,Mori17} can be used to form the density matrix, $|E\rangle\langle E|$. If ETH holds, then 
\begin{align}
\langle E|\mathcal{O}|E\rangle={\rm tr}(\mathcal O e^{-\beta H})/Z,
\end{align}
up to exponentially small in $L$ corrections.

Another way to view this result is to consider the moving average of an observable. It is readily shown that in the absence of degeneracies this projects onto the diagonal ensemble~\cite{Kollar08,Rigol08} at late times: 
\begin{align}
\label{eq: diagonal ensemble}
\overline{\langle \mathcal{O}(t\to\infty)\rangle}\to \sum_E |c_E|^2 \langle E|\mathcal{O}|E\rangle.
\end{align}
If ETH holds,  $\langle E|\mathcal{O}|E\rangle$ is essentially the same in every eigenstate that has appreciable overlap $c_E$ with the initial state. The diagonal matrix elements of the observable can then be pulled out of the sum over energy and the remaining sum becomes one, independent of the initial state, due to unitarity.

When the many-body spectrum acquires exact degeneracies, the diagonal ensemble is no longer directly applicable. One must first diagonalize the observable in the basis of degenerate eigenstates, and only then use the diagonal ensemble. When the distribution of eigenvalues of arbitrary local operators in a degenerate manifold becomes sharply peaked at the thermal expectation value (with the width decreasing with increasing system size), those eigenstates satisfy ETH. 

\begin{figure*}[t!]
\flushleft(a)\includegraphics[width=0.90\columnwidth]{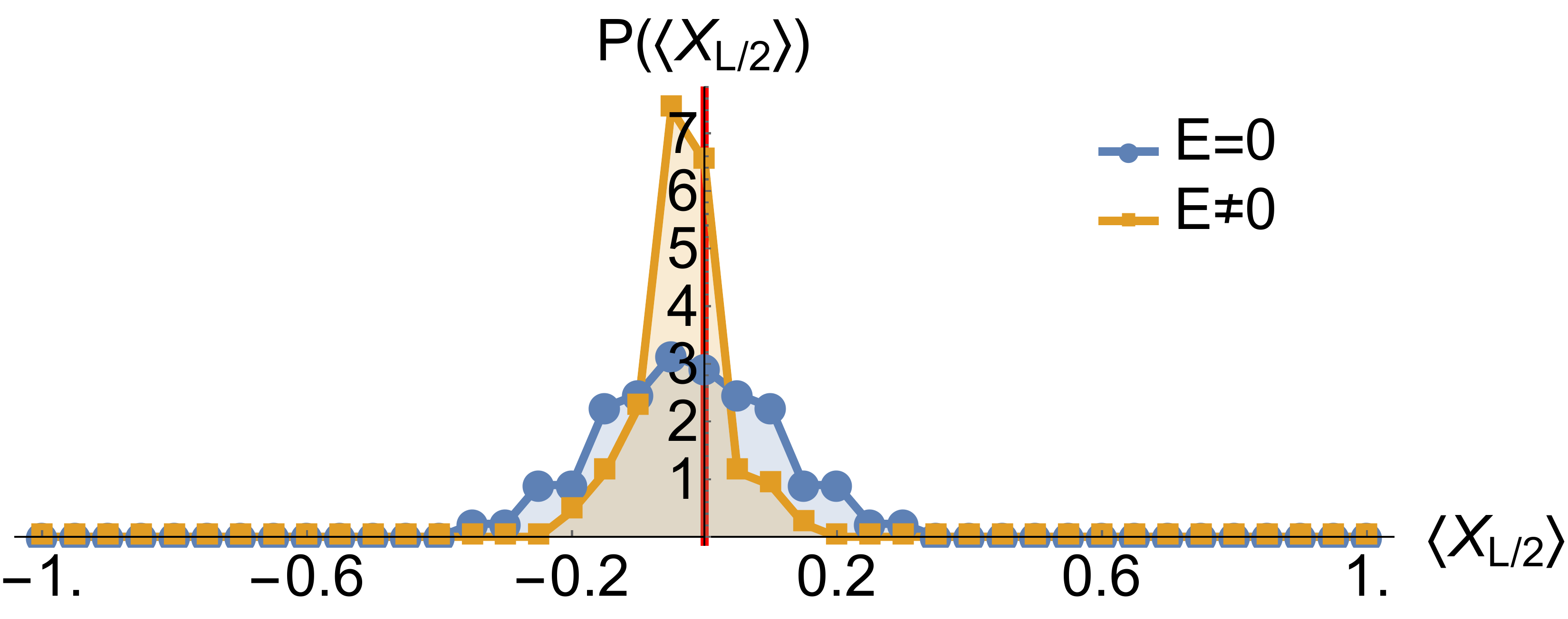}
\hfill
(d)\includegraphics[width=0.90\columnwidth]{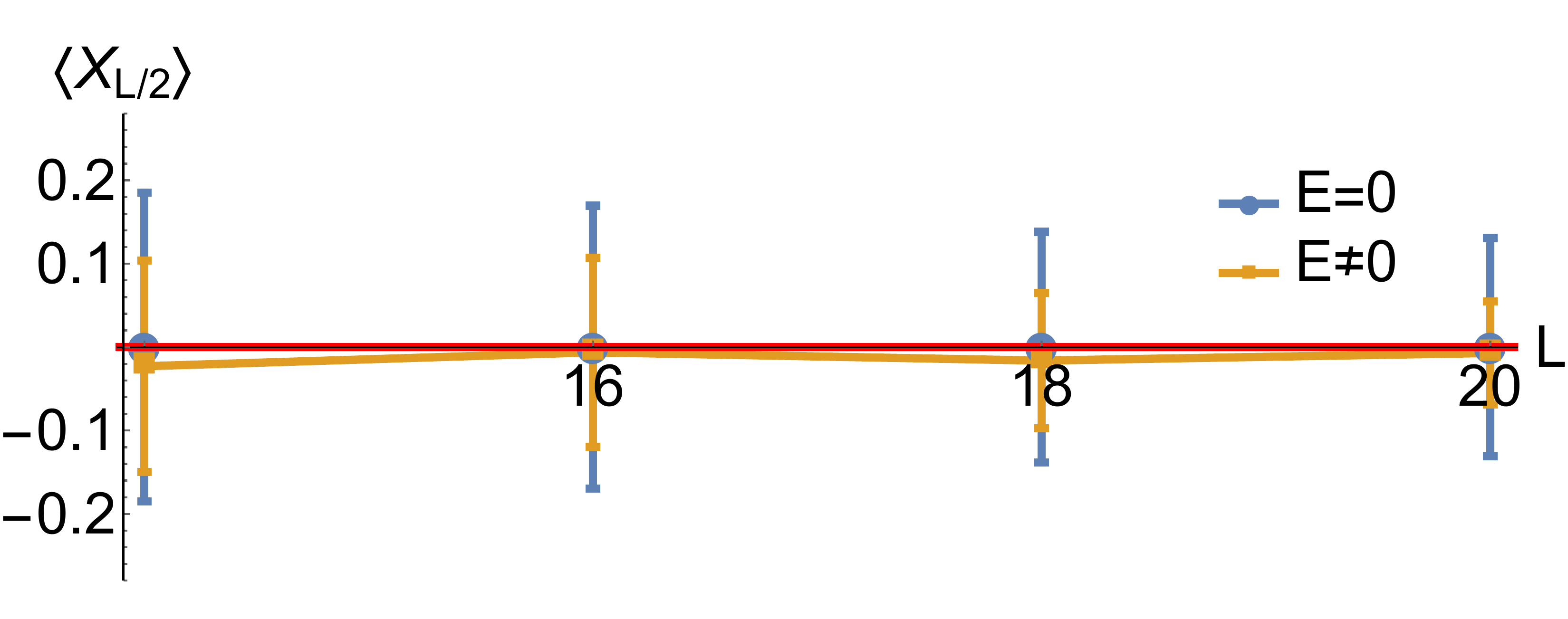}
\\
\flushleft(b)\includegraphics[width=0.90\columnwidth]{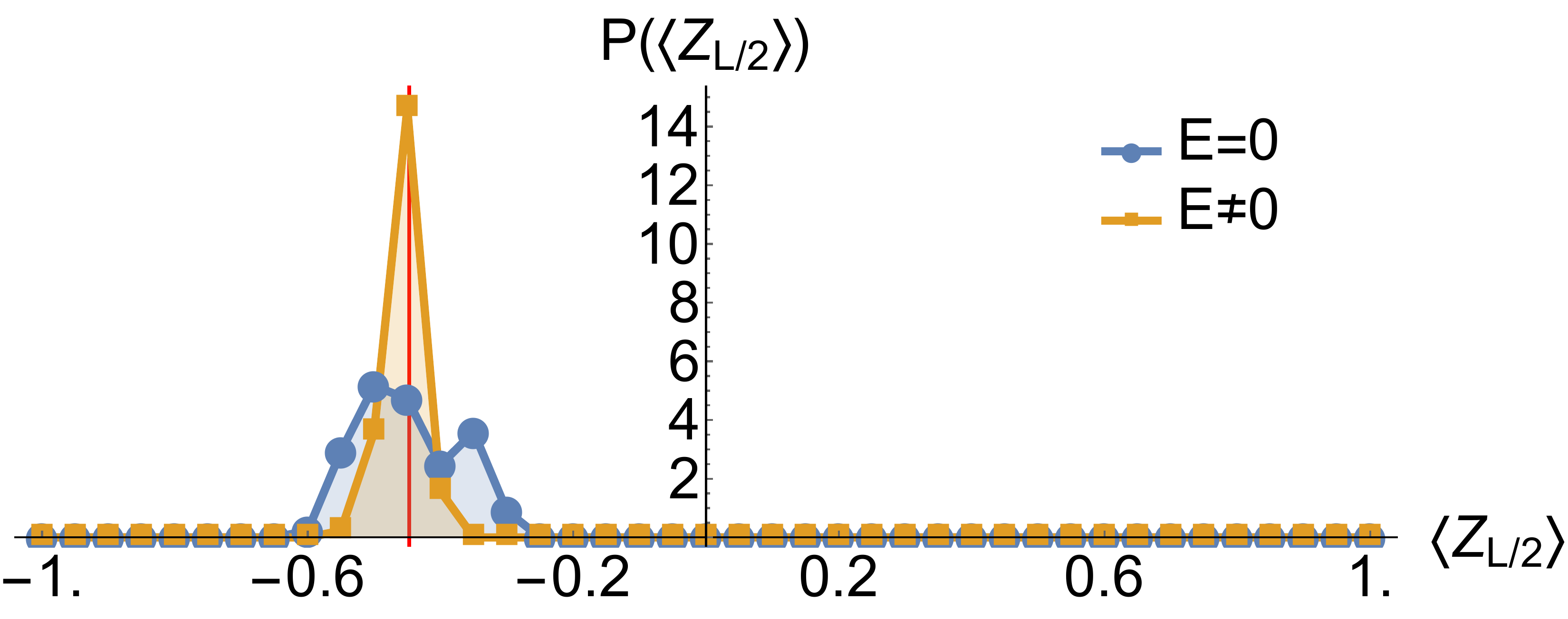}
\hfill
(e)\includegraphics[width=0.90\columnwidth]{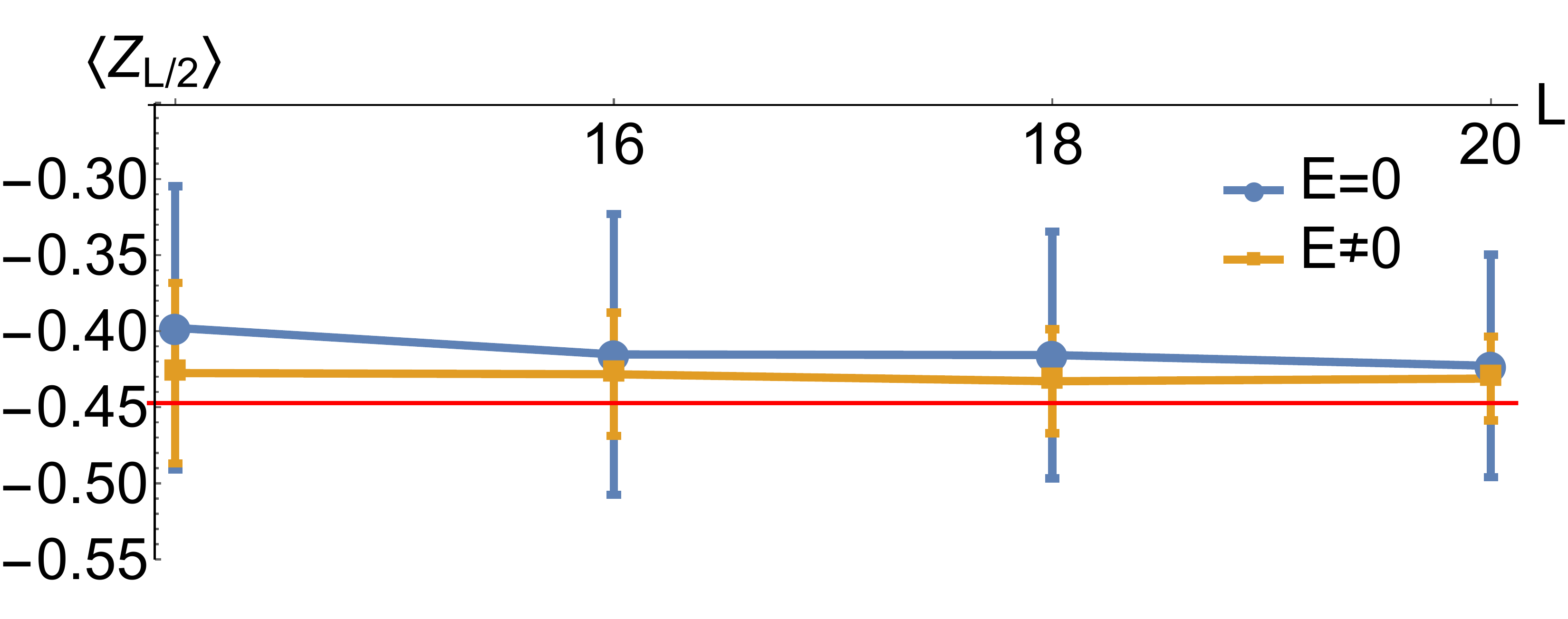}
\\
\flushleft(c)\includegraphics[width=1.0\columnwidth]{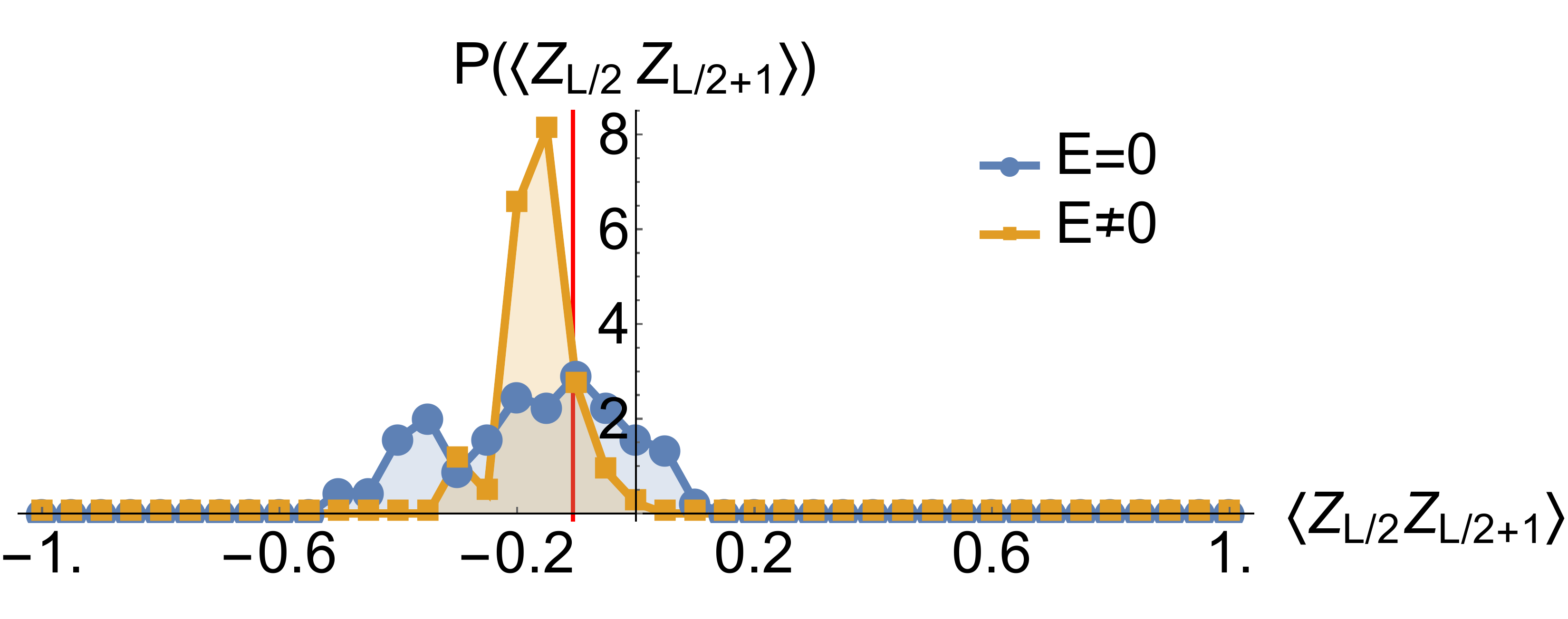}
\hfill
(f)\includegraphics[width=0.90\columnwidth]{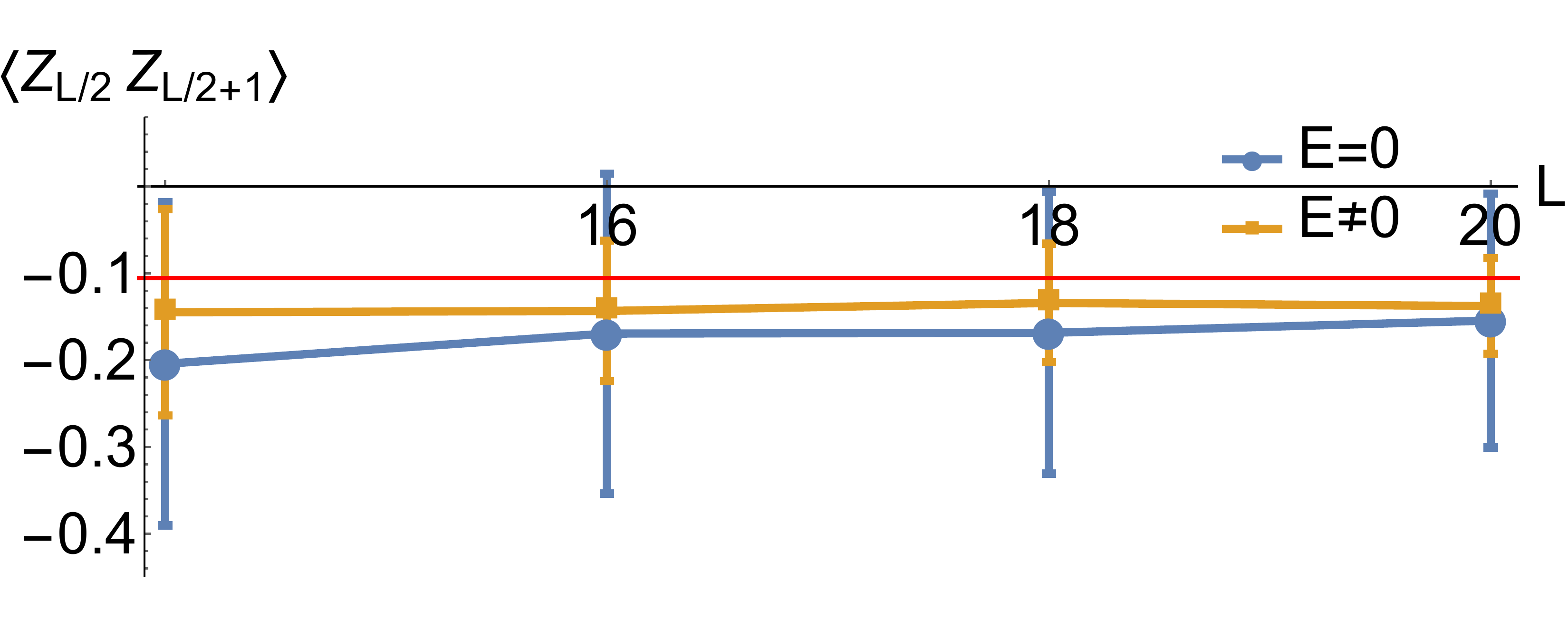}
\\
\caption{(Color online) 
Eigenstate thermalization within and outside the zero-mode manifold.  Panels (a)--(c) depict the distributions of diagonal matrix elements of the local operators (a) $X_{L/2}$, (b) $Z_{L/2}$, and (c) $Z_{L/2}Z_{L/2+1}$ over $N_{0}$ eigenstates at zero energy (blue) and at nonzero energy (yellow) at system size $L=20$.  For zero-energy states, the matrix elements are taken using linear combinations of the zero modes that diagonalize the operator in question.  The nonzero-energy states are chosen from an energy window centered around an energy density $E/L\sim1/200$, so that the zero- and nonzero-energy states have comparable effective temperatures.  In all cases, both the $E=0$ and $E\neq 0$ distributions are peaked near the infinite-system-size thermal values (red vertical lines) given by Eqs.~\eqref{eq:local_obs1}, \eqref{eq:local_obs2}, \eqref{eq:local_obs3}, for (a), (b), and (c) respectively. However, the $E=0$ distributions are significantly wider and less sharply peaked than the $E\neq 0$ distributions.  Panels (d)--(f)  depict the $L$-dependence of the average values computed from the distributions shown in (a)--(c), respectively.  The error bars indicate a region of uncertainty of one standard deviation above and below each data point.  The red lines indicate the thermal values given by Eqs.~\eqref{eq:local_obs1}, \eqref{eq:local_obs2}, and \eqref{eq:local_obs3}.  In all cases, both the $E=0$ and $E\neq 0$ values are consistent with the corresponding thermal values, and the magnitude of the error bars decreases as a function of $L$.  However, at any fixed $L$ the error bars on the $E=0$ value exceed those on the $E\neq 0$ value, indicating that the $E=0$ distribution is broader than the $E\neq0$ distribution.
}
\label{fig:local_obs}
\end{figure*}

We show the eigenvalue distribution of a few local observables in the manifold of zero modes of Eq.~\eqref{eq:H1} in Fig.~\ref{fig:local_obs} (a)--(c) (blue curves). As a function of system size $L$ the distributions become more strongly peaked near their infinite-temperature and infinite-size thermal expectation values, $\langle \mathcal{O}\rangle_0={\rm tr}(\mathcal{O})/\mathcal D$ [see Fig.~\ref{fig:local_obs} (d)--(f) (blue curves), where the size of the error bars decreases with system size]. This supports the notion that the states in the degenerate zero-mode manifold of the Hamiltonian \eqref{eq:H1} satisfy ETH. For the operators considered in Fig.~\ref{fig:local_obs} we find, for $L\to\infty$,
\begin{subequations}
\begin{align}
\label{eq:local_obs1}{\rm tr}(X_{L/2})/\mathcal D&=0,
\\
\label{eq:local_obs2}{\rm tr}(Z_{L/2})/\mathcal D&=1/\sqrt{5},
\\
\label{eq:local_obs3}{\rm tr}(Z_{L/2}Z_{L/2+1})/\mathcal D&=-1+2/\sqrt{5}.
\end{align} 
\end{subequations}
These values are indicated as red vertical lines in Fig.~\ref{fig:local_obs} (a)--(c) and red horizontal lines in Fig.~\ref{fig:local_obs} (d)--(f). We note that in Eqs.~\eqref{eq:local_obs2}-\eqref{eq:local_obs3} the non-vanishing value of the trace of the Pauli operators arises due to the constrained Hilbert space of Eq.~\eqref{eq:H1}.

It is interesting to compare the matrix-element distributions obtained from the zero-mode manifold to the distributions of diagonal matrix elements obtained from nearby nonzero-energy states.  The latter should also satisfy ETH at a temperature close to that of the zero modes (i.e.~near-infinite), as they are nondegenerate states in the middle of the many-body spectrum.  The results of this analysis are shown in Fig.~\ref{fig:local_obs} (a)--(c), where the diagonal-matrix-element distributions for nonzero energy states are shown in yellow.  At $L=20$, the latter distributions are significantly more sharply peaked about their mean value compared to the eigenvalue distributions in the zero-mode manifold, shown in blue.  Fig.~\ref{fig:local_obs} (d)--(f) compares the mean and standard deviation of the two distributions as a function of $L$.  Evidently, the nonzero-energy states yield a significantly more sharply peaked distribution at each system size studied. 

These results suggest an intriguing conclusion, namely that the states in the zero-mode manifold thermalize more ``slowly" as a function of $L$ than nearby states at similar energy densities.  It is likely, if ETH holds, that the observed discrepancy between the two sets of states will lessen in the thermodynamic limit.  Nevertheless, at the system sizes studied here, the difference between the widths of the zero- and nonzero-energy matrix-element distributions does not appear to diminish as a function of $L$.  We have also verified numerically using another model with exponentially many zero modes, namely the paramagnet \eqref{eq: H_para} perturbed by interactions of the form \eqref{eq: general deltaH}, that the same discrepancy in the two distributions arises.  Hence, it appears that this discrepancy is not a consequence of the constrained nature of the Fibonacci-chain model, but rather is a generic property of thermalizing quantum systems with zero modes.  One possible explanation for the ``slower" thermalization of the zero modes has to do with the fact that the zero-energy states have a constraint that their neighbors at nearby energies do not: a conserved chiral charge.  However, more work is necessary in order to sharpen these observations and identify the mechanism underlying the difference between the two distributions.

Despite these differences, in the thermodynamic limit, it is reasonable to expect both the $E=0$ and the $E\neq0$ diagonal-matrix-element distributions to become infinitely sharp and peaked at their thermal values.  If ETH holds, any eigenstate within or near the zero-energy manifold can be used to construct the microcanonical ensemble at that energy scale. As a result, ETH implies that the presence or absence of zero modes in the spectrum (even exponentially many) is irrelevant for the late-time dynamics of observables in the thermodynamic limit since the precise value of the energy of the eigenstate used is insignificant.

We stress that the validity of ETH depends both on the observable and on the initial state, and may be violated in certain circumstances where the dynamics displays non-ergodic behavior. One such example is the Fibonacci chain studied here, where the ``scarring" of the many-body wavefunctions leads to dynamics that are sensitive to the choice of initial state~\cite{Turner17}. However, further work is required to determine to what degree ETH is violated in this system. Another class of systems known to violate ETH are those which are many-body localized~\cite{Nandkishore15,Altman15}, where strong disorder precludes ergodic dynamics.

\section{Conclusion}\label{sec:conclusion}

In this paper, we have shown that an exponential number of protected many-body zero modes can arise in a large class of nonintegrable quantum spin chains with spectral-reflection and point-group symmetries. We showed that their robustness is guaranteed by an index theorem, and that they can be measured in systems that are relevant to several ongoing experiments \cite{Bernien17,Guardado-Sanchez17,Lienhard17}.  We have provided numerical evidence supporting the eigenstate thermalization of the manifold of zero modes, despite the fact that there is manifestly no level repulsion. Understanding the character and role of zero modes in the presence of (symmetry-preserving) disorder in the MBL regime where the breakdown of ETH occurs is an interesting direction for future work.

\acknowledgements{We thank Alexey Gorshkov and Timothy Hsieh for discussions, and Anushya Chandran and Vedika Khemani for sharing unpublished results. We acknowledge support from the Laboratory for Physical Sciences and Microsoft.  T.I. acknowledges a JQI postdoctoral fellowship.}

\appendix

\section{Sampling the Zero Mode Count}

\renewcommand\thefigure{\thesection\arabic{figure}}    
\setcounter{figure}{0}

\begin{figure}[b!]
\includegraphics[width=\columnwidth]{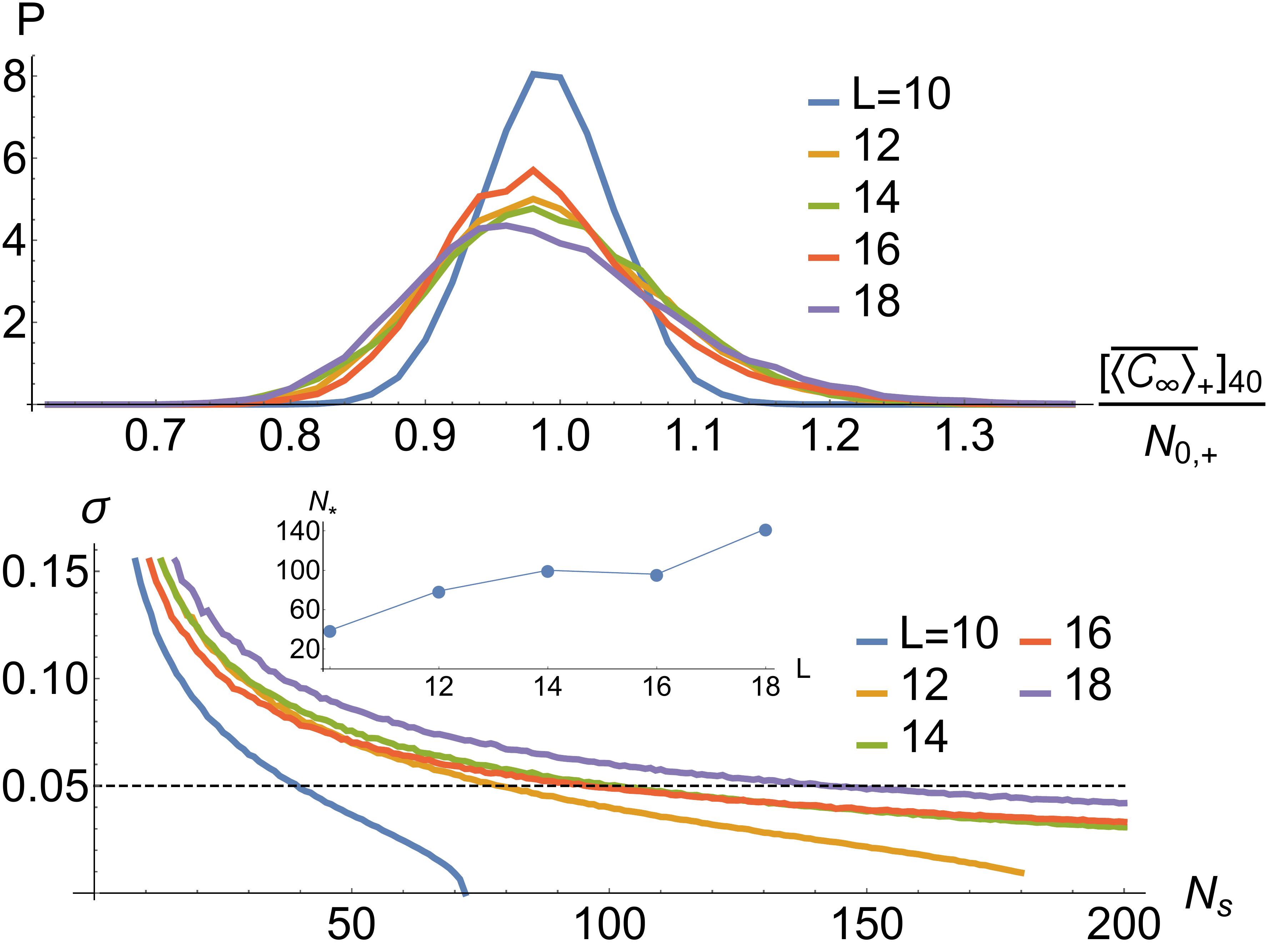}
\caption{(Color online)
Upper panel: Probability distribution $P$ of $[\overline{\langle\mathcal C_{\infty}\rangle}_{+}]_{40}$, c.f. Eq.~\eqref{eq:cinf_sampled}, for various system sizes at $\Delta=\delta=0$. Here
$[\overline{\langle\mathcal C_{\infty}\rangle}_{+}]_{40}$ is normalized against the number $N^{\,}_{0,+}$
of zero modes in the $\mathcal C=1$ sector.
20000 realizations of the random sample are used to generate each distribution. 
Lower panel: Standard deviation $\sigma$ of the distribution
$P$ as a function of the sample size $N^{\,}_{s}$.
The dashed line indicates a $5\%$ precision threshold.
The inset shows the sample size $N_{*}$ required to
reach this threshold as a function of $L$.}
\label{fig:PNs40_distributions}
\end{figure}

In this Appendix we show that it is possible to measure the quantity $\overline{N}_0$, Eq.~\eqref{eq:N0}, with reasonable accuracy by a random sampling of $\overline{\langle\mathcal{C}_\infty\rangle}_\psi$ over initial states $\psi$. Instead of summing over all initial states with fixed chiral charge, let us choose a random sample $s$ of $N_s$ initial states with fixed chiral charge $\pm1$.
For a fixed $N_s$, we can then define the quantity
\begin{equation}
\label{eq:cinf_sampled}
[\overline{\langle\mathcal C_{\infty}\rangle}_{\pm}]_{N_s}=\left|\frac{\mathcal D_{\pm}}{N_s}\sum_{\psi\in s} \overline{\langle\mathcal{C}_\infty\rangle}_\psi\right|,
\end{equation}
where $\mathcal D_{\pm}=1/2\, [\mathcal D(L) \mp (-1)^L\, a(L)]$ is the number of (constrained) $Z_i$ product states with $\mathcal C=\pm 1$, as an approximation to $|W_{\pm}|$. This approximation becomes exact when $N_{s}=\mathcal D_{\pm}$. More precisely, we can consider the \emph{distribution} $P$ of $[\overline{\langle\mathcal C_{\infty}\rangle}_{\pm}]_{N_s}$ over different realizations of the random sample $s$. As $N_{s}$ increases, the mean of $P$ approaches the number of zero modes with $\mathcal C=\pm 1$, $N_{0,\pm}$, while the standard deviation $\sigma$ of $P$ approaches zero.  Examples of the distribution $P$ for $\mathcal C=+1$, $N_s=40$, and varying system sizes are shown in the upper panel of Fig.~\ref{fig:PNs40_distributions}.
The lower panel of Fig.~\ref{fig:PNs40_distributions} depicts the decrease of $\sigma$ with $N_s$
as a function of system size.

What sample size $N_s$ is necessary to estimate $N_{0,\pm}$ to a fixed degree of precision? For a precision threshold of $5\%\, (\sigma=0.05)$, we find numerically that the number of required random samples $N_*$ scales much more slowly with system size than $\mathcal D_+\sim \mathcal D(L)/2$ over the range $10\leq L\leq 18$, as shown in the inset of Fig.~\ref{fig:PNs40_distributions}. This is an enormous simplification relative to the naive implementation of the protocol described before Eq.~\eqref{eq:obs}, which requires the preparation of every possible initial product state. For example, $\mathcal D_+=3383$ for $L=18$, but a random sample of only 150 of these initial states suffices to achieve the $5\%$ threshold.  Augmenting the naive protocol with these sampling techniques may render it feasible in the experimental setups of, e.g., Refs.~\cite{Bernien17, Guardado-Sanchez17,Lienhard17}.

\bibliography{refs_zero}

\end{document}